\documentclass[11pt,preprint2]{aastex}

\shorttitle{Grids of Atmospheres and Spectra}
\shortauthors{Kirby}

\begin{document}
\newcommand{\teff}{$T_{\rm{eff}}$}
\newcommand{\mathteff}{T_{\rm eff}}
\newcommand{\logg}{$\log g$}
\newcommand{\mathlogg}{\log g}
\newcommand{\afe}{[$\alpha$/Fe]}
\newcommand{\mathafe}{{\rm [\alpha/Fe]}}
\newcommand{\ah}{[$\alpha$/H]}
\newcommand{\mathah}{{\rm [\alpha/H]}}
\newcommand{\vt}{$v_t$}
\newcommand{\mathvt}{v_t}

\title{Grids of ATLAS9 Model Atmospheres and MOOG Synthetic Spectra}

\author{Evan~N.~Kirby\altaffilmark{1,2}}

\altaffiltext{1}{Hubble Fellow.}
\altaffiltext{2}{California Institute of Technology, 1200
  E.\ California Blvd., MC 249-17, Pasadena, CA 91125}

\keywords{stars:atmospheres}


\begin{abstract}

A grid of ATLAS9 model atmospheres has been computed, spanning
$3500~{\rm K} \le \mathteff \le 8000~\rm{K}$, $0.0 \le \mathlogg \le
5.0$, $-4.0 \le \mathrm{[M/H]} \le 0.0$, and $-0.8 \le \mathafe \le
+1.2$.  These parameters are appropriate for stars in the red giant
branch, subgiant branch, and the lower main sequence.  The main
difference from a previous, similar grid \citep{cas03} is the range of
\afe\ values.  A grid of synthetic spectra, calculated from the model
atmospheres, is also presented.  The fluxes are computed every
0.02~\AA\ from 6300~\AA\ to 9100~\AA.  The microturbulent velocity is
given by a relation to the surface gravity.  This relation is
appropriate for red giants, but not for subgiants or dwarfs.
Therefore, caution is urged for the synthetic spectra with $\log g >
3.5$ or for any star that is not a red giant.  Both the model
atmosphere and synthetic spectrum grids are available online through
VizieR.  Applications of these grids include abundance analysis for
large samples of stellar spectra and constructing composite spectra
for stellar populations.

\end{abstract}


\section{Introduction}
\label{sec:intro}

One approach to measuring the composition of a star from low- or
medium-resolution spectroscopy is to synthesize the stellar spectrum.
An essential ingredient in the synthesis is a model atmosphere.
Typically, a model atmosphere tabulates quantities such as density,
temperature, pressure, electron number density, and opacity as a
function of optical depth.  A grid of synthetic spectra at a range of
atmospheric parameters (effective temperature, surface gravity, and
composition) streamlines the measurement of these parameters from the
observed spectrum.  Therefore, it is useful to have a grid of model
atmospheres, from which a grid of synthetic spectra may be computed.

\citet{mih65} and \citet{str65} generated the first grids of non-gray,
continuum model atmospheres.  The use of opacity distribution
functions \citep[ODFs,][]{str66} greatly simplifies the computation of
such grids.  ODFs treat the absorption coefficient as smoothly varying
within wavelength subdivisions of the spectrum.  In effect, the
information about line opacity is compressed into a compact function,
which may be used to compute model atmospheres.

The grid of model atmospheres most relevant to the current work is
that of \citet{cas03}.  They computed a grid of model atmospheres with
the ATLAS9 program \citep{kur93a}.  \citet{kur70} described the
original ATLAS code in great detail.  \citeauthor{cas03} sampled a
wide range of effective temperatures (\teff), surface gravities
(\logg), and metallicity ([M/H]).  They also sampled two values of
\afe: 0.0 and +0.4.

One major use of stellar atmospheres is the computation of synthetic
stellar spectra.  Because it is computationally expensive to generate
a model atmosphere, it is sometimes wise to invest the computational
resources up-front by generating a grid of synthetic spectra.  The red
to near-infrared spectral region is the subject of this article.
There is a precedent for synthetic spectral grids in this spectral
region \citep[e.g.,][]{coe05,mun05,pal10}.  The synthetic spectral
grid here is unique because of its wide range in the $\alpha$
enhancement (\afe).

\begin{deluxetable}{lcccc}
\tablecolumns{5}
\tablewidth{0pt}
\tablecaption{Atmospheric Parameter Grid\label{tab:grid}}
\tablehead{\colhead{Quantity} & \colhead{Minimum} & \colhead{Maximum} & \colhead{Step} & \colhead{Number}}
\startdata
\teff\ (K) & $3500$ & $8000$ & $\left\{\begin{array}{lc}
                                               100 & \mathteff \le 5500 \\
                                               200 & \mathteff \ge 5600
                                               \end{array}\right.$ & 34 \\
\logg\ (cm~s$^{-2}$), $\mathteff \le 6800$ & 0.0 & 5.0 & 0.5 & 11 \\
\logg\ (cm~s$^{-2}$), $\mathteff \ge 7000$ & 0.5 & 5.0 & 0.5 & 10 \\
\protect{[M/H]} (atmospheres) & $-4.0$ & $0.0$ & $0.5$ & 9 \\
\protect{[M/H]} (spectra) & $-4.0$ & $0.0$ & $0.1$ & 41 \\
\afe & $-0.8$ & $+1.2$ & $0.1$ & 21 \\
\enddata
\tablecomments{Sections~\ref{sec:atmospheres} and \ref{sec:spectra}
  give the parameterization of microturbulent velocity, $\xi$.}
\end{deluxetable}

\citet{kir10} measured the detailed abundances of thousands of red
giant stars in the dwarf satellite galaxies of the Milky Way.  In
order to do so, they required a grid of synthetic spectra.  A grid of
model atmospheres was computed at a range of \afe\ so that the
compositions of the atmospheres would be consistent with the spectra.
The inspiration for this procedure was the grid of \citet{cas03}.
However, the presence in dwarf galaxies of stars with large
\afe\ \citep{fre10} and small \afe\ \citep{let10} required a broader
range of \afe\ in the model atmospheres.  Therefore,
\citeauthor{cas03}'s computation of model atmospheres was repeated,
but for a wide range in \afe.  A grid of synthetic spectra was
computed based on the grid of model atmospheres.

The purpose of this article is to make these grids publicly available.
Those with possible interests in these grids include researchers who
want to measure the elemental abundances for a large number of far-red
spectra or who want to generate composite spectra of stellar
populations for an integrated light analysis.  Other uses are also
encouraged.

Table~\ref{tab:grid} gives the details of the grid parameters.  The
step size for [M/H] is smaller for the spectra than for the
atmospheres (see Section~\ref{sec:spectra}).
Section~\ref{sec:atmospheres} describes the generation of the grid of
model atmospheres, and Section~\ref{sec:spectra} describes the grid of
synthetic spectra.  Section~\ref{sec:summary} explains how the data
may be retrieved from the VizieR online catalog.


\section{ATLAS9 Model Atmospheres}
\label{sec:atmospheres}

\begin{figure*}[t!]
\centering
\includegraphics[width=0.95\textwidth]{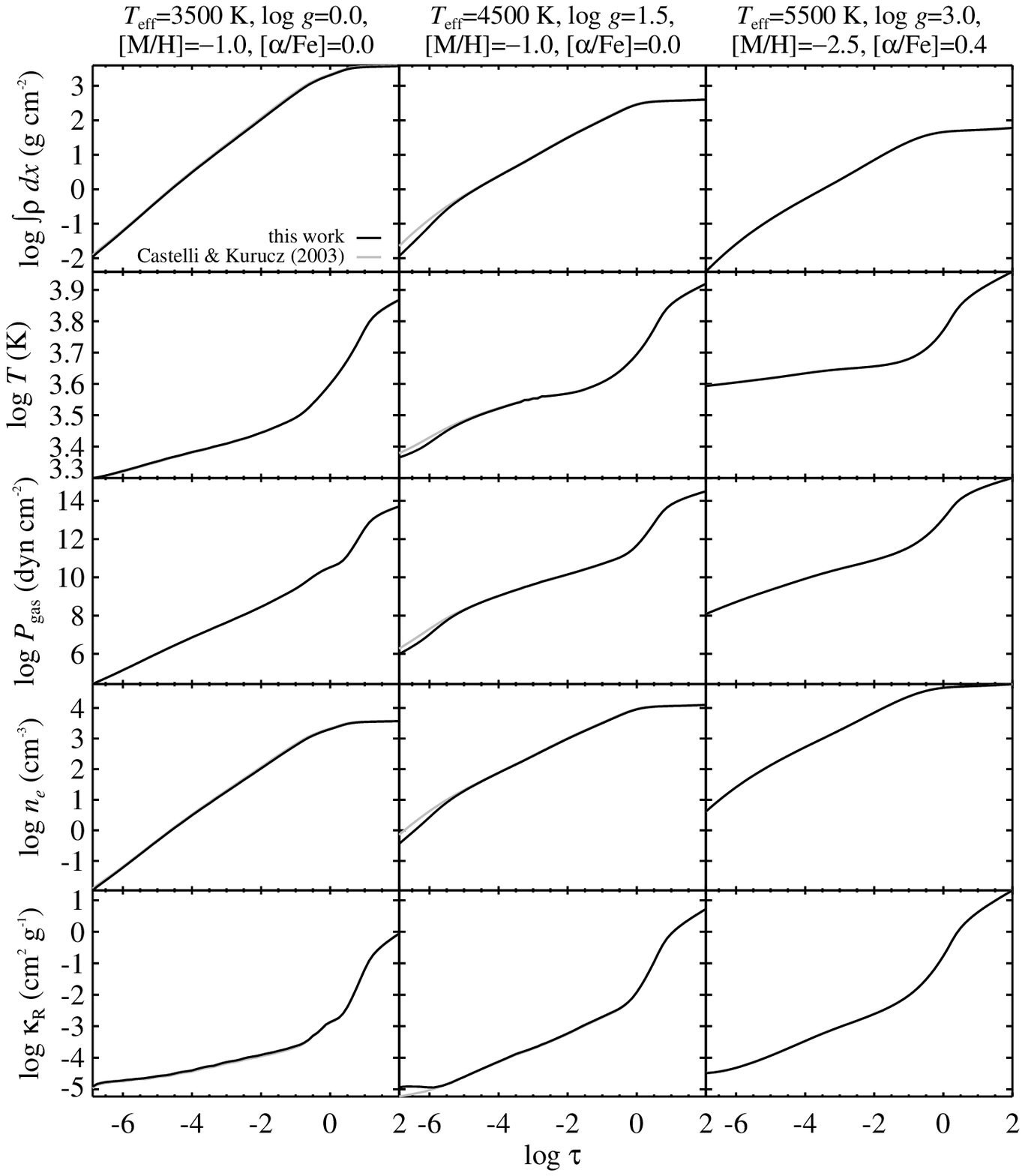}
\caption{ATLAS9 model atmospheres computed here ({\it black}) compared
  to the ATLAS9 model atmospheres with the same parameters computed by
  \citet[][{\it gray}]{cas03}.  The black and gray lines overlap for
  much of this figure.  Each column shows a different star whose
  parameters are given at the top of the column.  The five rows show
  the mass depth ($\int \rho\, dx$), temperature ($T$), gas pressure
  ($P_{\rm gas}$), electron number density ($n_e$), and Rosseland mean
  opacity ($\kappa_{\rm R}$) as a function of the logarithm of optical
  depth ($\log \tau$).\label{fig:atmospheres}}
\end{figure*}

The grid (Table~\ref{tab:grid}) of plane-parallel model atmospheres in
local thermodynamic equilibrium (LTE) was computed with
\citeauthor{kur93a}'s (\citeyear{kur93a}) DFSYNTHE, KAPPA9, and ATLAS9
codes ported into Linux \citep{sbo04,sbo05} and compiled with the
Intel Fortran compiler 11.1.  All computations were performed in
serial on 256 processors at the University of California Santa Cruz
Pleiades Linux computing cluster.  The compositions of the ODFs were
the scaled solar abundances of \citet{and89} except that the abundance
of iron was $12 + \log \epsilon (\mathrm{Fe}) =
7.52$\footnote{$\epsilon({\rm X})$ is the ratio of the number density
  ($n$) of atoms of element X to the number density of hydrogen atoms:
  $\epsilon({\rm X}) = n({\rm X})/n({\rm H})$.  However, note that the
  ATLAS code accepts abundances expressed as $n({\rm X})/n_{\rm
    total}$.  This work assumes the solar helium-to-hydrogen number
  ratio of 0.098 \citep{and89}.  Also note that the solar abundance
  pattern assumed here \citep{and89} is different from the solar
  abundance pattern of \citet{gre98}, which is the default for the
  DFSYNTHE and ATLAS9 codes.}.  The $\alpha$ elements O, Ne, Mg, Si,
S, Ar, Ca, and Ti were additionally scaled by the parameter \afe.

The starting point for an ATLAS9 model atmosphere is an ODF.  ODFs
were computed in a manner identical to \citet{cas03} except for the
elemental composition.  In summary, DFSYNTHE \citep[described
  by][]{cas05} was used to generate an ODF at each point in the
stellar parameter grid.  F.~Castelli's scripts
(\url{http://wwwuser.oat.ts.astro.it/castelli/}) were employed for
this purpose.  These scripts compute ODFs at lower and higher
effective temperatures than the grid presented here, but keeping the
file formats consistent with Castelli's scripts simplified later steps
in computing the model atmospheres.  The lists of atomic and some
molecular (including H$_2$, HD, CH, C$_2$, CN, CO, NH, OH, MgH, SiH,
and SiO but not including TiO or H$_2$O) transitions were those of
R.~L. Kurucz, downloaded at
\url{http://wwwuser.oat.ts.astro.it/castelli/sources/dfsynthe.html}.
See \citet{kur90} for details of the line lists.  The molecular line
lists for TiO and H$_2$O, downloaded at the same URL, were provided by
\citet{sch98} and \citet{par97}, respectively.  The KAPPA9 code in the
same software package as DFSYNTHE computed Rosseland mean opacities
from the ODFs.

For consistency with \citeauthor{cas03}'s (\citeyear{cas03}) grid of
ATLAS9 atmospheres, convective overshooting was turned off, and the
mixing length parameter for convection was $l/H_p = 1.25$.  For each
combination of \teff, \logg, [M/H], and \afe, there are two
atmospheres with two different values of microturbulent velocity,
$\xi$.  The values are the two velocities among 0, 1, 2, and
4~km~s$^{-1}$ that bracket the microturbulent velocity appropriate for
the star's surface gravity (Equation~2 of \cite{kir09}):
$\xi~(\mathrm{km~s}^{-1}) = 2.13 - 0.23\,\log g$.

ATLAS9 requires an input model atmosphere as an initial guess.  An
ATLAS9 model atmosphere with similar \teff, \logg, and [M/H] from
\citeauthor{cas03}'s grid was used for this purpose.  At least 30
iterations were computed for each stellar atmosphere.  After 30
iterations, convergence was tested by examining the difference in the
flux and flux derivative between the last two iterations.
Infrequently, these differences exceeded 1\% for the flux or 10\% for
the flux derivative.  In these cases, a different initial model
atmosphere was used as input for 30 new ATLAS9 iterations.  If the
resulting atmosphere was still not converged, additional iterations
were computed until the convergence criteria were satisfied.

In very rare cases, it was not possible to achieve these criteria with
any number of iterations.  These atmospheres were left as is.  Most of
these atmospheres corresponded to very luminous, warm red giants
($6000~\mathrm{K} \le \mathteff \le 8000~\mathrm{K}$ and $0.0 \le
\mathlogg \le 0.5$).  These stars lose mass in winds, which violate
ATLAS9's assumption of hydrostatic equilibrium.  Stars with these
atmospheric parameters are exceptionally rare.  Therefore, convergence
problems in this region of the grid are unlikely to cause problems for
real astrophysical applications.  The model atmospheres for M dwarfs
($\mathteff \la 4000$~K, $\mathlogg \la 3.5$) also violated the
convergence criteria to a small degree in the outer layers, especially
at $\rm{[Fe/H]} \la -1$.  M dwarf atmospheres have low electron
densities and non-blackbody spectral energy distributions, which cause
them to be out of statistical equilibrium \citep{sch99}.  The outer
layers affect the strongest absorption lines, which are often formed
out of LTE even for warmer and larger stars.

Figure~\ref{fig:atmospheres} shows some example stellar atmospheres.
The parameters for each atmosphere were chosen to lie near a 14~Gyr
isochrone from the Victoria-Regina set of isochrones \citep{van06}.
For all three models, the microturbulent velocity is $\xi = 2~{\rm
  km~s}^{-1}$.  The parameters plotted are those relevant to the
generation of synthetic spectra.  From top to bottom, they are the
logarithms of the mass depth ($\int \rho\, dx$), temperature ($T$),
gas pressure ($P_{\rm gas}$), electron number density ($n_e$), and
Rosseland mean opacity ($\kappa_{\rm R}$).  The independent variable
is the logarithm of the optical depth.  The \citet{cas03} models for
identical stellar parameters are also shown in gray for comparison.
Subtle differences between the two sets of atmospheres---most visible
in the middle column of Figure~\ref{fig:atmospheres}---occur due to
the slight differences in elemental composition and different criteria
for convergence.  The largest differences occur in the shallowest
layers of the atmosphere, which are important only for the formation
of the strongest saturated lines.

\section{MOOG Synthetic Spectra}
\label{sec:spectra}

\begin{figure*}[t!]
\centering
\includegraphics[width=\textwidth]{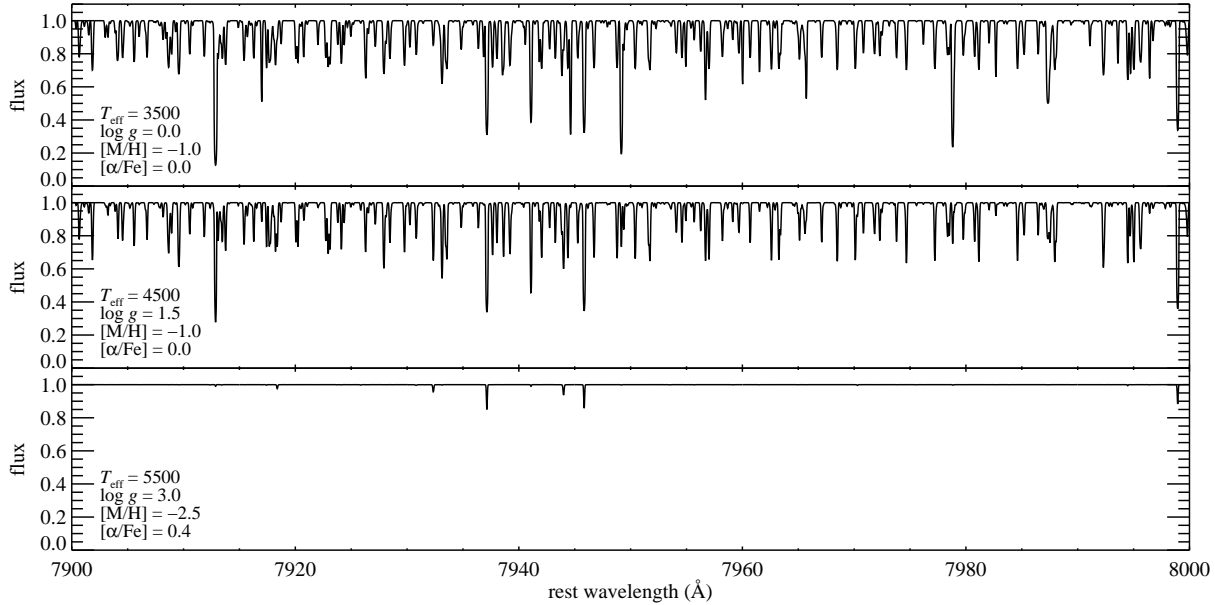}
\caption{Examples of three synthetic spectra computed with MOOG.  The
  stellar parameters and model atmospheres are identical to those
  shown in Figure~\ref{fig:atmospheres}.\label{fig:synths}}
\end{figure*}

\begin{figure*}[t!]
\centering
\includegraphics[width=\textwidth]{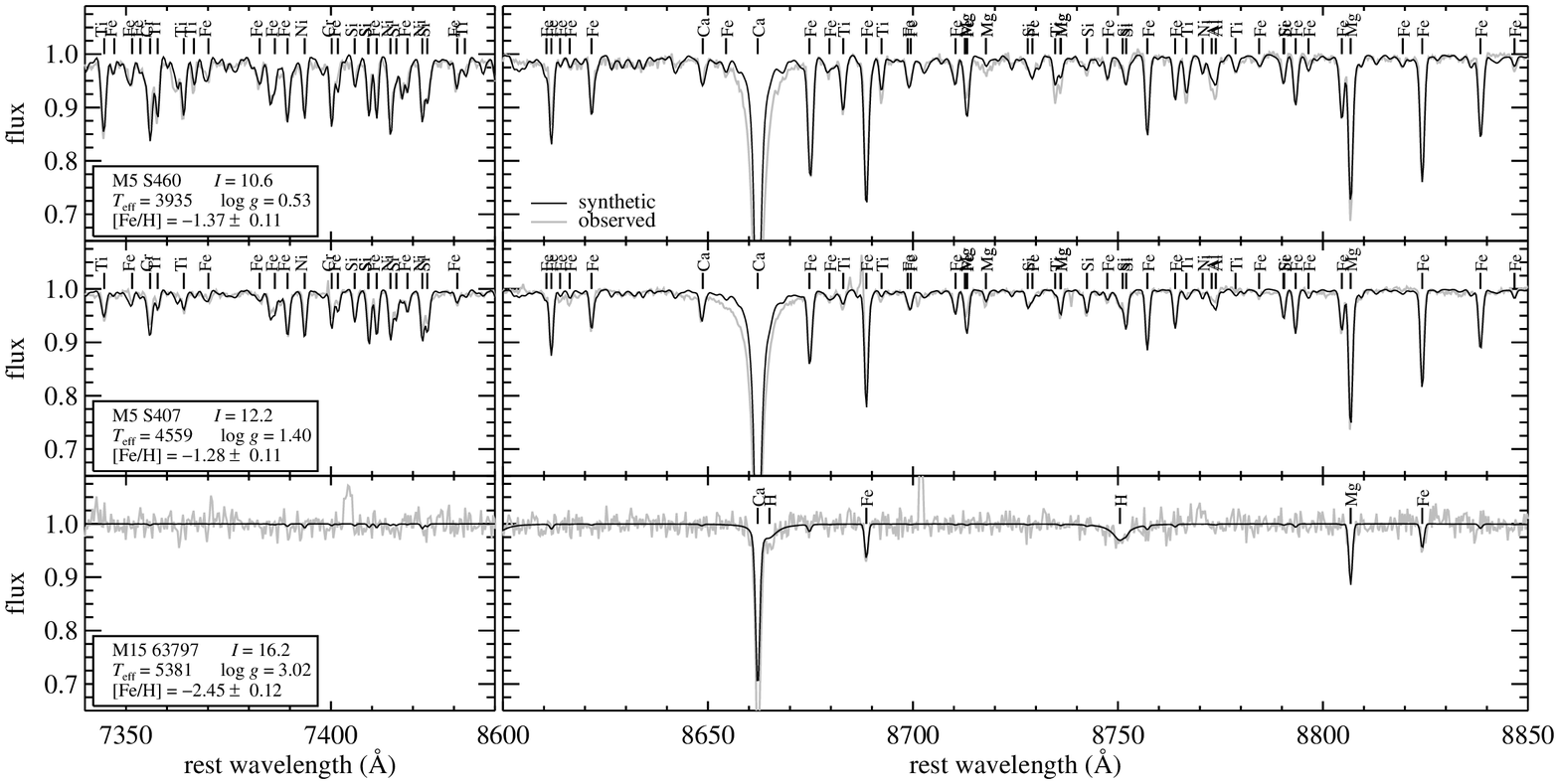}
\caption{Examples of three synthetic spectra ({\it black}) compared to
  spectra observed with DEIMOS ({\it gray}).  The synthetic spectra
  were interpolated in the grid in order to replicate the atmospheric
  parameters given in the legend at the left of each panel.  The
  interpolated spectra were smoothed to match the resolution of the
  observed spectra ($R \approx 6500$).  Some of the prominent stellar
  absorption lines are labeled with the abbreviation of the
  responsible element.\label{fig:spectra}}
\end{figure*}

The grid of synthetic spectra was computed with the 2007 version of
the plane-parallel, LTE code MOOG \citep{sne73}.  The code was
modified to run in parallel via OpenMP on 256 processors on the
Pleiades cluster.  The parallelization was trivial, with different
regions of the grid being computed simultaneously.  Each spectrum was
computed on a single processor.  MOOG was also modified to output only
the synthetic spectrum into an unformatted binary file.  All other
output was suppressed to save disk space and file access time.  The
normalized flux is computed every 0.02~\AA\ from 6300~\AA\ to
9100~\AA.  The units of the spectrum are such that the continuum is
unity.  The continuum shape is not computed.  One spectrum is computed
at each grid point.

MOOG requires two data tables as input: a model atmosphere and a line
list.  The model atmospheres were provided by the ATLAS9 atmospheres
described in Section~\ref{sec:atmospheres}.  However, the spectral
grid was five times finer in the [M/H] dimension than the model
atmospheres (see Table~\ref{tab:grid}).  Therefore, intermediate
atmospheres spaced at 0.1~dex in [M/H] were generated by linear
interpolation of the physical variables (not the logarithms of the
physical values).

The microturbulent velocity chosen for each synthetic spectrum is a
value appropriate for red giants (Equation~2 of \citeauthor{kir09}
\citeyear{kir09}): $\xi~(\mathrm{km~s}^{-1}) = 2.13 - 0.23\,\log g$.
The atmosphere used to generate a spectrum was a linear interpolation
between the two bracketing values of $\xi$.  This linear interpolation
was performed in conjunction with linear interpolation in [M/H], where
necessary.

\citet*{kir08} gave the complete line list, but the version used here
has been modified according to \citet{kir09}.  The original line list
contained atomic lines from the Vienna Atomic Line Database
\citep[VALD,][]{kup99}, molecular lines from \citet{kur92}, and
hyperfine atomic lines from \citet{kur93b}.  Oscillator strengths were
modified to match observed line strengths in Arcturus and the Sun.
\citet{kir09} replaced the oscillator strengths ($\log gf$) for some
\ion{Fe}{1} lines with the values of \citet{fuh06}.  For those
\ion{Fe}{1} lines not shared with \citet{fuh06}, 0.13~dex was
subtracted from $\log gf$.  The only molecules in the line list are
CN, C$_2$, and MgH.  The opacity for almost every line is considered
only within 1~\AA\ of the line's central wavelength.  Some lines are
considered at every wavelength in the spectrum.  These are H$\alpha$,
the hydrogen Paschen series redward of 8370~\AA, the \ion{Ca}{2}
triplet at 8498, 8542, and 8662~\AA, and \ion{Mg}{1} $\lambda 8807$.

Figure~\ref{fig:synths} shows 100~\AA\ (7900--8000~\AA) of the
synthetic spectra computed from the model atmospheres shown in
Figure~\ref{fig:atmospheres}.  The full spectral range is 63 times
what is shown.  The abundances used to compute the line strengths are
completely consistent with the abundances used to compute the model
atmospheres.

Figure~\ref{fig:spectra} shows synthetic spectra compared to spectra
observed with the Keck/DEIMOS medium-resolution spectrograph.
\citet{kir10} obtained these spectra.  These stars were chosen to have
atmospheric parameters fairly close to those shown in
Figures~\ref{fig:atmospheres} and \ref{fig:synths}.  The stars are red
giants in the globular clusters M5 and M15.  The atmospheric
parameters for the synthetic spectra are given in the legend at the
left of each panel \citep[measured by][]{kir10}.  The synthetic
spectra were linearly interpolated, pixel by pixel, to reproduce the
measured atmospheric parameters.  Finally, the synthetic spectral
resolution has been degraded to match the DEIMOS instrumental
resolution.  In the spectrograph configuration used, profiles of
unresolved lines have $\rm{FWHM} \approx 1.2$~\AA\ almost independent
of wavelength.  Therefore, the resolving power is $R \approx 6200$ at
7400~\AA\ and $R \approx 6700$ at 8700~\AA.

The grid of synthetic spectra has several shortcomings, some of which
are enumerated here.

\begin{enumerate}

\item The only molecules included are CN, C$_2$, and MgH.  The number
  of TiO lines makes including them in the spectra computationally
  prohibitive.  Therefore, cool, metal-rich stars are not modeled well
  in this grid.  In reality, their spectra have TiO features that
  become extremely strong as the temperature decreases and metallicity
  increases.

\item Strong lines are not modeled well, mostly due to non-LTE
  effects.  For example, Figure~\ref{fig:spectra} clearly shows that
  the observed wings and core of the \ion{Ca}{2} $\lambda 8662$ line
  are much stronger than in the synthetic spectrum.

\item Although a relation between microturbulent velocity and surface
  gravity has been established, not all stars obey this trend.  In
  fact, the relation that \citet{kir09} adopted was based only on red
  giants.  The relation is not appropriate for most subgiants and
  dwarfs.  The user is strongly cautioned against using any synthetic
  spectra with $\log g > 3.5$ without verifying that the
  microturbulent velocity is appropriate for the specific star under
  consideration.

\end{enumerate}



\section{Online Data Access}
\label{sec:summary}

The grids of model atmospheres and synthetic spectra may be obtained
at the VizieR online catalogue service \citep{och00} at the following
URL: \url{http://vizier.u-strasbg.fr/viz-bin/VizieR?-source=VI/134}.
The user has two options to access the catalog.  First, the catalog
may be accessed either through the VizieR online interface, which
includes an option to plot the synthetic spectra in a web browser.
Second, one or both of the grids may be downloaded via FTP or HTTP.
The compressed file sizes are 310~MB for the atmospheres and 46~GB for
the spectra.

\acknowledgments The author thanks Joel Primack for sharing his time
on the Pleiades Astrophysics Computing Cluster at the University of
California Santa Cruz, supported by an NSF-MRI grant; Judy Cohen for
sharing a list of helpful references; Fran\c{c}ois Ochsenbein for his
help in preparing the data for online access via VizieR; F.~Castelli
for her helpful comments; and R.~Kurucz, the referee, whose
suggestions improved this article.  Support for this work was provided
by NASA through Hubble Fellowship grant 51256.01 awarded to ENK by the
Space Telescope Science Institute, which is operated by the
Association of Universities for Research in Astronomy, Inc., for NASA,
under contract NAS 5-26555.  This research has made use of the VizieR
catalogue access tool, CDS, Strasbourg, France.

\end{document}